\documentclass {pspum-l}
\usepackage{amsthm}
\usepackage{amsmath}
\usepackage{amssymb}
\usepackage{graphicx}
\usepackage{bm}

\begin{document}

\title {Nodal domains on graphs - How to count them and why?}
\author {Ram Band$^\natural$, Idan Oren$^\natural$ and Uzy Smilansky$^{\natural,\S}$}
\address {$\natural$  Department of Physics of Complex Systems, The Weizmann
Institute of Science, Rehovot 76100, Israel.}
\address {$\S$ Cardiff School of Mathematics and WIMCS,
Cardiff University, Senghennydd Road, Cardiff CF24 4AG, UK}

\thanks{MSC classes:  34B45, 05C50, 15A18}
\thanks{The work was supported by the Minerva Center for non-linear Physics
and the Einstein (Minerva) Center at the Weizmann Institute, and by
grants from the EPSRC (grant 531174), GIF (grant I-808-228.14/2003),
and BSF (grant 2006065).}

\begin{abstract}The purpose of the present manuscript is to collect
known results and present some new ones relating to nodal domains on
graphs, with special emphasize on nodal counts. Several methods for
counting nodal domains will be presented, and their relevance as a
tool in spectral analysis will be discussed.
\end{abstract}

\maketitle

\renewcommand{\baselinestretch}{1.5}

\newtheorem*{thm3.1}{Theorem 3.1}
\newtheorem*{thm3.2}{Theorem 3.2}
\newtheorem*{thm3.3}{Theorem 3.3}

\newtheorem*{lem4.1}{Lemma 4.1}
\newtheorem*{thm4.2}{Theorem 4.2}

\newtheorem*{thm5.1}{Theorem 5.1}
\newtheorem*{thm5.2}{Theorem 5.2}

\section{Introduction}

Spectral graph theory deals with the spectrum and the eigenfunctions
of the Laplace operator defined on graphs. The study of the
eigenfunctions, and in particular, their nodal domains  is an
exciting and rapidly developing research direction. It is an
extension to graphs of the investigations of nodal domains on
manifolds, which started already in the 19th century by the
pioneering work of Chladni on the nodal structures of vibrating
plates. Counting nodal domains started with Sturm's oscillation
theorem which states that a vibrating string is divided into exactly
\textit{n} nodal intervals by the zeros of its $\textit{n}^{th}$
vibrational mode. In an attempt to generalize Sturm's theorem to
manifolds in more than one dimension, Courant formulated his nodal
domains theorem for vibrating membranes, which bounds the number of
nodal domains of the $\textit{n}^{th}$ eigenfunction by \textit{n}
\cite{Cou53}. Pleijel has shown later that Courant's bound can be
realized only for finitely many eigenfunctions \cite{Ple56}. The
study of nodal domains counts was revived after Blum \emph{et al}
have shown that nodal count statistics can be used as a criterion
for quantum chaos \cite{BGS02}. A subsequent paper by Bogomolny and
Schmit illuminated another fascinating connection between nodal
statistics and percolation theory \cite{Bog02}. A recent paper by
Nazarov and Sodin addresses the counting of nodal domains of
eigenfunctions of the Laplacian on  $\mathbb{S}^2$ \cite {Sod07}.
They prove that on average, the number of nodal domains increases
linearly with $n$, and the variance about the mean is bounded. At
the same time, it was shown that the nodal sequence - the sequence
of numbers of nodal domains ordered by the corresponding spectral
parameters - stores geometrical information about the domain \cite
{GPS06} . Moreover, there is a growing body of numerical and
theoretical evidence which shows that the nodal sequence can be used
to distinguish between isospectral manifolds. \cite {GSS05,BKP07}.

In the present paper we shall focus on the study of nodal domains
on graphs, and show to what extent it goes hand in hand or
complements the corresponding results obtained for Laplacians on
manifolds.

The  paper is designed as follows: The next chapter summarizes some
elementary  definitions and background material necessary to keep
this paper self contained. Next, we survey the known results
regarding counting nodal domains on graphs and state a new theorem
regarding the morphology of nodal domains. After these
preliminaries, we present a few counting methods of nodal domains on
graphs. Finally, the intimate connection between nodal sequences and
isospectrality on graphs will be reviewed, and some open problems
will be formulated.

\section{Definitions, notations and background}

 A \textit{graph} $\mathcal{G}=(\mathcal{V},\mathcal{B})$ is a set
of vertices $\mathcal{V}=\{1,2,\ \cdots  V \}$ of size
$V\equiv|\mathcal{V}|$ and a set of undirected bonds (edges)
$\mathcal{B}$ of size $B\equiv|\mathcal{B}|$, such that $\{i,j\}\in
\mathcal{B}$ if the vertices $i$ and $j$ are connected by a bond. In
this case we say that vertices $i$ and $j$ are \textit{adjacent} and
denote this by $i\sim j$. The \textit{degree} (\textit{valency}) of
a vertex is the number of bonds which are connected to it.  A graph
is called $v$-regular if all its vertices are of degree $v$.
Throughout the article, and unless otherwise stated, we deal with
connected graphs with no multiple bonds or loops (a bond which
connects a vertex to itself). A well known fact in graph theory is
that the number of independent cycles in a graph, denoted by $r$ is
equal to:
\begin{equation}
r=B-V+Co \label{eq:cycles}
\end{equation}
where $Co$ is the number of connected components in $\mathcal{G}$.
We note that $r$ is also the rank of the fundamental group of the
graph.  A \textit{tree} is a graph for which $r=0$. \\
Let $g$ be a subgraph of $\mathcal{G}$. We define the
\emph{interior} of $g$ as the set of vertices whose adjacent
vertices are also in $g$. The \emph{boundary} of $g$ is the set of
vertices in $g$ which are not in its interior.

A graph $\mathcal{G}$ is said to be  \textit{properly colored} if
each vertex is colored so that adjacent vertices have different
colors. $\mathcal{G}$ is \textit{k-colorable} if it can be properly
colored using \textit{k} colors. The \textit{chromatic number}
$\chi(\mathcal{G})$ is \textit{k} if $\mathcal{G}$ is
\textit{k}-colorable and not (\textit{k}-1)-colorable. A very simple
observation, which we will use later, is that $\chi(\mathcal{G})\leq
V$.

 $\mathcal{G}$ is called \textit{bipartite} if its chromatic number is 1 or
2. However, since a chromatic number 1 corresponds to a graph with
no bonds, and we are dealing only with connected graphs, we can
exclude this trivial case and say that for a bipartite graph,
$\chi=2$. The vertex set of a bipartite graph $\mathcal{G}$ can be
partitioned into two disjoint sets, say $\mathcal{V}_{1}$ and
$\mathcal{V}_{2}$, in such a way that every bond of $\mathcal{G}$
connects a vertex from $\mathcal{V}_{1}$ with a vertex from
$\mathcal{V}_{2}$. We then have the following notation:
$\mathcal{G}=(\mathcal{V}_{1}\cup
\mathcal{V}_{2},\mathcal{B})$ \cite{CDS79}.\\

The \textit{adjacency (connectivity) matrix} of $\mathcal{G}$ is the
symmetric $V\times V$ matrix $C=C(\mathcal{G})$ whose entries are
given by:
\begin{displaymath}
C_{ij}=\left\{ \begin{array}{ll} 1, & \textrm{if $i$ and $j$ are
adjacent}\\
0, & \textrm{otherwise}
\end{array} \right.
\end{displaymath}

Laplacians on graphs can be defined in various ways. The most
elementary way relies only on the topology (connectivity) of the
graph, and the resulting Laplacian is an operator on a discrete and
finite-dimensional Hilbert space. These operators or their
generalizations to be introduced below will be referred to as
``discrete" or ``combinatorial" Laplacians. One can construct the
Laplacian operator as a differential operator if the bonds are
endowed with a metric, and appropriate boundary conditions are
required at the vertices. The resulting operator should be referred
to as the ``metric'' Laplacian. However, because the metric
Laplacian is identical with the free Schr\"odinger operator (i.e.
with no potential) on the graph, one often refers to this system as
a ``Quantum Graph'' - a misnomer which is now hard to eradicate. In
the sequel we shall properly define and discuss the relevant
versions of Laplacians on graphs.

The \textit{discrete}  Laplacian, of $\mathcal{G}$, is the matrix
 \begin{equation}
 L(\mathcal{G})=D-C\ ,
 \label {eq:dislap}
 \end{equation}
where $D$ is the diagonal matrix whose $i^{th}$ diagonal entry is
the degree of the vertex $i$, and $C$ is the adjacency matrix of
$\mathcal{G}$. A \textit{generalized} Laplacian, $L^{'}$ is a
symmetric $V\times V$ matrix with off-diagonal elements defined by:
$L^{'}_{ij}<0$ if vertices $i$ and $j$ are adjacent, and
$L^{'}_{ij}=0$ otherwise. There are no constraints on the diagonal
elements of $L^{'}$.

The eigenvalues of $L(\mathcal{G})$ together with their
multiplicities, are known as the \textit{spectrum} of $\mathcal{G}$.
To each eigenvalue corresponds (at least one) eigenvector whose
entries are labeled by the vertices indexes. It is well known that
the eigenvalues of the combinatorial Laplacian are non-negative.
Zero is always an eigenvalue and its multiplicity is equal to the
number of connected components of $\mathcal{G}$. An important
property regarding spectra of large $v$-regular graphs is that the
limiting spectral distribution is symmetric about $\lambda=v$, and
is supported on the interval $[v-2\sqrt{v-1},v+2\sqrt{v-1}]$
\cite{M81}.

An extensive survey of the spectral theory of discrete Laplacians
can be found in \cite{Mo91,Chung, BiLeSt07}. \\

To define \textit{quantum graphs} a metric is associated to
$\mathcal{G}$. That is, each bond is assigned a positive length:
$L_b \in (0,\infty)$.  The coordinate along the bond $b$ is
denoted by $x_b$. The total length of the graph will be denoted by
$\mathcal{L}=\sum_{b\in \mathcal{B}} L_b$. This enables to define
the metric Laplacian (or free Schr\"odinger operator) on the graph
as the negative second derivative $-\frac{{\rm d}^2 \ }{{\rm d}
x^2}$ on each bond. The domain of this operator on the graph is
the space of functions which connect continuously across vertices
and which belong to the Sobolev space $W^{2,2}(b)$ on each bond b.
Moreover, vertex  boundary conditions are imposed to render the
operator self adjoint. We shall consider in this paper the Neumann
and Dirichlet boundary conditions:
\begin{eqnarray}
{\rm Neumann \ condition \ on \ the \ vertex\ } i :\ \ & & \ \ \
\sum_{b\in S^{(i)}}  \left. \frac{{\rm d}\ \ \ }{{\rm d} x_{b} }\
\psi_{b}(x_{b})\right |_{x_{b}=0} \ = \ 0\ \ ,  \label{eq:Neumann boundary_cond} \\
{\rm Dirichlet \ condition \ on \ the \ vertex\ } i :\ \ & &   \ \ \
\ \ \ \ \ \ \left. \psi_{b}(x_{b})\right |_{x_{b}=0} =0\ ,
 \label{eq:Dirichlet boundary_cond}
\end{eqnarray}
where $S^{(i)}$ denotes the group of bonds which emerge from the
vertex $i$ and the derivatives in (\ref{eq:Neumann boundary_cond})
are directed out of the vertex $i$. The eigenfunctions are the
solutions of the bond Schr\"odinger equations:
\begin{eqnarray}
\ \ & & \ \ \forall b\in \mathcal{B} \  \ -\frac{{\rm d}^2 \ }{{\rm
d} x^2} \psi_b=k^2\psi_b ,
 \label{eq:schrodinger_eq}
\end{eqnarray}
which agree on the vertices and satisfy at each vertex boundary
conditions of the type (\ref{eq:Neumann boundary_cond}) or
(\ref{eq:Dirichlet boundary_cond}). The spectrum
$\{k_n^2\}_{n=1}^{\infty}$ is discrete, non-negative and unbounded.
One can generalize the metric Laplacian by including potential and
magnetic flux that are defined on the bonds. Other forms of boundary
conditions can also be used. However, these generalizations will not
be addressed here, and the interested reader is referred to two
recent reviews \cite{GS06,Kuch04}.

Finally, two graphs, $\mathcal{G}$ and $\mathcal{H}$, are said to be
\textit{isospectral} if they posses the same spectrum (same
eigenvalues with the same multiplicities).  This definition holds
both for discrete and quantum graphs.

\section{Nodal domains on graphs}
\emph{Nodal domains} on graphs are defined differently for
discrete and metric graphs. \\
$\bullet \ ${\emph{Discrete graphs:}} Let
$\mathcal{G}=(\mathcal{V},\mathcal{B})$ be a graph and let $\bf
f$=$(f_{1},f_{2},\ldots,f_{V})$ be a real vector. We associate the
real numbers $f_{i}$ to the vertices of $\mathcal{G}$ with
$i=1,2,\ldots,V$. A \textit{nodal domain} is a maximally connected
subgraph of $\mathcal{G}$ such that all vertices have the same sign
with respect to $\bf f$. The number of nodal domains with respect to
a vector $\bf f$ is called a \textit{nodal domains count}, and will
be denoted by $\nu(\bf f)$. The maximal number of nodal domains
which can be achieved by a graph $\mathcal{G}$ will be denoted by
$\nu_{\mathcal{G}}$. The \textit{nodal sequence} of a graph is the
number of nodal domains of eigenvectors of the Laplacian, arranged
by increasing eigenvalues.  This sequence will be denoted by $\{
\nu_n \}_{n=1}^{V}$.

\noindent The definition of nodal domains should be sharpened if we
allow zero entries in $\bf f$. Two definitions are then natural:

\begin {itemize}
\item { \emph{A strong positive (negative) nodal domain}} is a
maximally connected subgraph $\mathcal{H}$ of $\mathcal{G}$ such
that $\textit{f}_i>0$ ($\textit{f}_i<0$) for all $i\in \mathcal{H}$.
\item {\emph{A weak positive (negative) nodal domain}} is a maximally
connected subgraph $\mathcal{H}$ of $\mathcal{G}$ such that
$\textit{f}_i\geq0$ ($\textit{f}_i \leq0$) for all $i\in
\mathcal{H}$.
\end {itemize}
In both cases, a positive (negative) nodal domain must consist of
at least one positive (negative) vertex. According to these
definitions, it is clear that the weak nodal domains count is
always smaller or equal to the strong one. \\
$\bullet \ ${\emph{Metric graphs:}} Nodal domains are connected
domains of the metric graph where the eigenfunction has a constant
sign. The nodal domains of the eigenfunctions are of two types.
The ones that are confined to a single bond are rather trivial.
Their length is exactly half a wavelength and their number is on
average $\frac {k \mathcal{L}}{\pi}$. The nodal domains which
extend over several bonds emanating from a single vertex vary in
length and their existence is the reason why counting nodal
domains on graphs is not a trivial task. The number of nodal
domains of a certain eigenfunction on a general graph can be
written as
\begin{equation}
\hspace{-5mm}  \mu_n =\frac{1}{2}\sum_{i}\sum_{b\in S^{(i)}} \left
\{
  \left\lfloor\frac{k_n L_{ b }}{\pi}\right\rfloor+\frac{1}{2} \left( 1-(-1)^{
      \lfloor\frac{k_n L_{b}}{\pi} \rfloor}
    \mathrm{sign}[\phi_i]\mathrm{sign}[\phi_j]
  \right )\right  \}  -B +V \
  \label{eq:NNB2}
\end{equation}
where $\lfloor x \rfloor$ stands for the largest integer which is
smaller than $x$, and $\phi_i, \phi_j$ are the values of the
eigenfunction at the vertices connected by the bond $b=\{i,j\}$
\cite{GSW04}. (\ref{eq:NNB2}) holds for the case of an
eigenfunction which does not vanish on any vertex: $\forall i
\,\,\,\phi_i\neq0$, and there is no cycle of the graph on which
the eigenfunction has a constant sign.  The last requirement is
true for high enough eigenvalues where half the wavelength is
smaller than the length of the shortest bond. This restriction,
which is important for low eigenvalues, was not stated in
\cite{GSW04}.

Nodal domains on quantum graphs can be also defined and counted in
an alternative way.  Given an eigenfunction, we can associate to it
the vector ${\bm{\phi}}=(\phi_1, \ldots, \phi_V)$ of its values on
the vertices and count the nodal domains of this vector as in the
case of a discrete graph, explained above. The reasoning behind this
way of counting is that the values of the eigenfunction on the
vertices $\{\phi_i\}_{i=1}^{V}$ together with the eigenvalue $k^2$
store the complete information about the values of the eigenfunction
everywhere on the graph.  We thus have two independent ways to
define and count nodal domains on metric graphs. To distinguish
between them we shall refer to the first as {\it metric} nodal
domains, and the number of metric domains in the $n^{th}$
eigenfunction will be denoted by $\mu_n$. The domains defined in
terms of the values of the eigenfunction on the vertices will be
referred to as the {\it discrete} nodal domains. The number of the
discrete nodal domains in the $n^{th}$ eigenfunction will be denoted
by $\nu_n$, similar to the notation of this count for the discrete
graphs.

As far as counting nodal domains is concerned, trees behave as one
dimensional manifolds, and the analogue of Sturm's oscillation
theory applies for the eigenfunctions of the discrete \cite{Bi03}
and the metric Laplacians \cite{AlO92,PPAO96,PP04,Sch06}, as long
as the eigenvector (or the eigenfunction) does not vanish at any
vertex. Thus we have $\nu_n=n$ for discrete tree graphs and
$\mu_n=n$ for metric ones.

Similarly, Courant's theorem applies for the eigenfunctions of both
the discrete and the metric versions of the Laplacian on any graph:
$\nu_n\leq n$, \, $\mu_n\leq n$, \cite{DGLS01,GSW04}. It should be
noted that there is a correction due to multiplicity of the $n^{th}$
eigenvalue and the upper bound becomes $n + m - 1$, where $m$ is the
multiplicity \cite{DGLS01}. However, sharper lower and upper bounds
for the number of nodal domains were discovered recently. Berkolaiko
provided a lower bound for the nodal domains count for both the
discrete and the metric cases \cite{Be06}. He showed that the nodal
domains count of the $n^{th}$ eigenfunction of the Laplacian (either
discrete or metric) has no less than $n-r$ nodal domains ($r$ is the
number of independent cycles in the graph). Again, this is valid if
the eigenfunction has no zero entries and it belongs to a simple
eigenvalue. When $n-r<0$, this result is trivial since a nodal
domains count is positive by definition.  We note that for metric
graphs this theorem does not hold when the discrete count is used.
This can be explained by the simple observation that $n-r$ grows
unbounded while the discrete count is bounded by the number of
vertices.

A global upper bound for the nodal domains count of a graph
$\mathcal{G}$ was derived in \cite{Or07}: The maximal number of
nodal domains on $\mathcal{G}$ was proven to be smaller or equal to
$\nu_{\mathcal{G}}V-\chi+2$, where $\chi$ is the chromatic number of
$\mathcal{G}$. This bound is valid for any vector, not only for
Laplacian eigenvectors.

To end this section we shall formulate and prove a few results which
show that not all possible subgraphs can be nodal domains of
eigenvectors of the discrete Laplacians of $v$-regular graphs. The
topology and connectivity of nodal domains are restricted, and the
restrictions depend on whether the eigenvalue is larger or smaller
than the spectral mid-point $v$.
\begin{thm3.1}
\label{Pleijel-like theorems for graphs}
 Let $\mathcal{G}$ be a
$v$-regular graph. Then the following statements hold:

\noindent \textbf{i}. For all eigenvectors with eigenvalue $\lambda
> v$ the nodal domains do not have interior vertices.

\noindent \textbf{ii}. For all eigenvectors with eigenvalue $\lambda
<v$, all the nodal domains consist of at least two vertices.

\noindent \textbf{iii}. For all eigenvectors with eigenvalue
$\lambda < v-k$ (and $k<v$), in every nodal domain there exists at
least one vertex with a degree (valency) which is larger than $k$.
\end{thm3.1}

\begin{proof}[\textbf{Proof}]

Let $\bf{f}$ be an eigenvector with no zero entries of the discrete
Laplacian, corresponding to an eigenvalue $\lambda$. Let $g$ be a
nodal domain of $\mathrm{\bf{f}}$.

\noindent \textbf{i}. Assume that $i$ is an interior vertex in $g$.
Hence, the signs of $f_j$ for all $j \sim i$ are the same as the
sign of $f_i$. This is not compatible with
\begin{equation}
-\sum_{j\sim i}f_j = (\lambda-v)f_i
\end{equation}
for $\lambda >v$. Hence $g$ cannot have any interior vertices.

\noindent \textbf{ii}. Assume that the subgraph $g$ consists of a
single vertex $i$. Thus on all its neighbors $j\sim i$, the sign of
$f_j$ is different from the sign of $f_i$. This is not compatible
with
\begin{equation}
-\sum_{j\sim i}f_j = -  (v- \lambda)f_i
\end{equation}
for $\lambda < v$. Hence $g$ cannot consist of a single vertex.

\noindent \textbf{iii}.  Denote  the complement of the nodal domain
(subgraph) $g$ in $\mathcal{V}$ by $g^c$. For all the vertices $i$
in  $g$
\begin{equation}
(L\textbf{f})_{i}\ =\ vf_{i}-\sum_{j\in g}C_{j,i} f_{j}-\sum_{l\in
g^c}C_{l,i} f_{l}\ =\ \lambda f_{i} \ .
\end{equation}
Summing over $i\in g$ we  get:
\begin{equation}
(v-\lambda)\sum_{i\in g}f_{i} =\sum_{i\in g} \left ( \sum_{j\in
g}C_{j,i} f_{j}+\sum_{l\in g^c}C_{l,i} f_{l}\right) \label{eq:proof
3.1(3)}
\end{equation}
Assuming for convenience that $f_i$ are positive for $i\in g$, the
rightmost sum in the equation above is non positive, and therefore
\begin{equation}
(v-\lambda)\sum_{i\in g}f_{i}\ \le \ \sum_{i\in g}  \sum_{j\in
g}C_{j,i} f_{j} = \sum_{i\in g}v_{i} f_{i}\ \le \ \hat{v} \sum_{i\in
g} f_{i}\ .  \label{eq:proof 3.1(4)}
\end{equation}
Here, $v_{i}= \sum_{j\in g} C_{j,i}$ is the valency (degree) of the
$i^{th}$ vertex in $g$, and $\hat{v}$ denotes the largest valency in
the subgraph. Since it is assumed that $\lambda < v-k$ we get
\begin{equation}
k\ < \ \hat{v} \ ,
\end{equation}
which completes the proof.
\end{proof}
This theorem holds also for an eigenvector $\bf{f}$ which has zero
elements with the only exception being the failure of part \emph{i}
when using the weak count. The case $\lambda=v$ deserves special
attention. As long as the nodal domain under study has no vanishing
entries, it cannot consist of a single vertex nor can it have
interior vertices. Namely, for $\lambda = v$, statements \emph{i}
and \emph{ii} of Theorem 3.1. are valid simultaneously. Otherwise,
one should treat separately the strong and the weak counts.  For the
strong count, and  $\lambda=v$ a nodal domain cannot have an
interior vertex. However, using the weak count for $\lambda=v$ one
finds that no single vertex domains can exists, as in Theorem
3.1.\emph{ii}.

Item \emph{iii} of Theorem 3.1 can be used to provide a $\lambda$
dependent bound on the number of nodal domains of eigenvectors
corresponding to eigenvalues $\lambda < v$. Define the integer $k$
as  $k= v- \lceil \lambda \rceil $. Theorem 3.1.\emph{iii} implies
that every nodal domain occupies at least $k+2$ vertices. Thus,
their number is bounded by $\frac{V}{k+2}$. Courant theorem
guarantees that the number of nodal domains is bounded by the
spectral count $\mathcal{N}(\lambda)$. This information together
with the known expression for the expectation value of
$\mathcal{N}(\lambda)$ over the ensemble of random graphs, enable
us to show that for large $v$ and $V$, the bound $\frac{V}{k+2}$
is more restrictive than the Courant bound. Unlike Pleijel's
result, this bound is not uniform for the entire spectrum, and it
applies only to the lower half of the eigenvalues with $\lambda
<v$.

Theorem 3.1 can be easily extended to the nodal properties of the
eigenvectors of the generalized Laplacian, provided that the weights
at each vertex sum up to a constant $v$ which is the same for the
entire graph.

\section{How to count nodal domains on graphs?}
When discussing nodal domains counting, we must make a clear
distinction between \emph{algorithmic} and \emph{analytic}
methods. In the first class, we include computer algorithms. They
vary in efficiency and reliability, but they have one feature in
common, namely, that the number of nodal domains is provided not
as a result of a computation, but rather, it follows from a
systematic counting process. The most widely used method is the
Hoshen-Kopelman algorithm (HK) for counting nodal domains on
2-dimensional domains \cite{HK76}. Analytic methods provide the
number of nodal domains as a functional of the function and the
domain under study. The functional might be quite complicated, and
not efficient when implemented numerically. An example of an
analytical method for nodal domains counting in one dimension, is
given by
\begin{equation}
\nu = \int_a ^b  \delta\left(f(x)\right)\left | \frac{{\rm d}
f(x)}{{\rm d}x}\right |{\rm d}x \ + \ 1 \ ,
\end{equation}
where the nodal domains of $f(x)$, in the interval  $[a,b]$ are
provided (assuming that $f(a)f(b)\ne 0$). While counting in 1-d is
simple, there is no analytic counting method for computing the
number of nodal domains in higher dimensions: the complicated
connectivity allowed in high dimensions renders the counting
operation too non local.


Graphs, which are in some sense intermediate between one and two
dimensions still allow several analytic counting methods which we
discuss here. An example of an analytical count is given by (\ref
{eq:NNB2}). The HK algorithm is well suited for graphs which are
grids. However, it is not as efficient when the graph under study
is highly irregular. Although the HK algorithm fails for very
complex graphs, other algorithms, called \textit{labeling
algorithms},
display linear efficiency (\cite{DST92},\cite{FG96}).\\
Method \emph{III.} in the following list, in addition of providing
an analytical expression for the nodal domain count, can also be
implemented as a computer algorithm. We show that it performs as
efficiently as the labeling algorithm.

The counting methods that we present here are aimed for the
discrete counting of both discrete and metric graphs. In what
follows, we assume that a vector $\bf f$ is associated to the
vertex set with entries $f_i$. The nodal domains are defined with
respect to $\bf f$.

\subsection{Method \emph{I.} - Counting nodal domains in terms of
flips}  \label{subsectionI}

We define a \textit{flip} as a bond on the graph which connects
vertices of opposite signs with respect to a vector $\bf f$. The
sign vector of $\bf f$, denoted by $\widetilde{\bf f}$, is defined
by  $\widetilde{f}_{i}\equiv \textrm{sign}(f_{i})$. For the time
being, it is assumed that $\bf f$ has no zero entries. The general
situation will be discussed later. We denote the set of flips on
the graph by $\mathcal{F}(\bf f)$:
\begin{equation}
\mathcal{F}(\textbf{f})=\{(u,v)\in \mathcal{B}\mid f_{v}f_{u}<0\}
\ .
\end{equation}
The cardinality of $\mathcal{F}(\bf f)$ will be denoted by
$\mathrm{F}(\bf f)$.

\begin{lem4.1}
The number of flips of a sign vector $\widetilde{\bf f}$, can be
expressed as
\begin{equation}
    \mathrm{F}(\bf{f})=\mathrm{F}(\widetilde{\bf f})=\frac{1}{4}(\widetilde{\bf f},L\widetilde{\bf f})
    \ .
    \label{quadratic form}
\end{equation}
\end{lem4.1}

\begin{proof}[\textbf{Proof}]
Using:

\begin{eqnarray}
(\widetilde{\bf f},L\widetilde{\bf f}) &=& \frac{1}{2}\sum_{v\sim
u}
(\widetilde{f}_{v}-\widetilde{f}_{u})^{2}\\
(\widetilde{f}_{v}-\widetilde{f}_{u})^{2} &=& \left\{
\begin{array}{ll} 4, & \textrm{if $\widetilde{f}_{v}$ and $\widetilde{f}_{u}$ have opposite signs}\\
0, & \textrm{if $\widetilde{f}_{v}$ and $\widetilde{f}_{u}$ have the
same sign}
\end{array} \right.
\end{eqnarray}

\end{proof}
Using the number of flips, one can get an expression for the number
of nodal domains:
\begin{thm4.2}
Given a connected graph $\mathcal{G}$ on $V$ vertices, $B$ bonds
(and $r$ cycles) and a vector $\bf f$, then the number of nodal
domains of $\bf f$ is:
\begin{equation}
\nu(\mathrm{\bf{f}})=\frac{1}{4}(\widetilde{\bf f},L\widetilde{\bf
f})+V-B+l(\mathrm{\bf{f}})=\frac{1}{4}(\widetilde{\bf
f},L\widetilde{\bf f})-(r-l(\mathrm{\bf{f}}))+1
    \label{nodal domains via flips}
\end{equation}
\end{thm4.2}
\noindent where $l(\bf{f})$ is the number of independent cycles in
$\mathcal{G}$ of constant sign (with respect to $\bf f$).  The
second equality above is based on equation (\ref{eq:cycles}).

\begin{proof}[\textbf{Proof}]
Let us remove all the flips from the graph. We are now left with a
possibly disconnected graph $\widetilde{\mathcal{G}}$. There is a
bijective mapping between components of $\widetilde{\mathcal{G}}$
and nodal domains of $\mathcal{G}$. Hence, the number of
components in $\widetilde{\mathcal{G}}$ is equal to the nodal
domains count of $\mathcal{G}$ with respect to $\bf f$. Let the
number of nodal domains in $\mathcal{G}$ be denoted by $\nu(\bf
f)$. Using (\ref{eq:cycles}), it is clear that for the $i^{th}$
component (where $i=1,2,\ldots,\nu(\bf{f})$):
$r_{i}=B_{i}-V_{i}+1$ where $r_{i}$, $B_{i}$ and $V_{i}$ are the
number of cycles, bonds and vertices of the $i^{th}$ component,
respectively. It is also clear that all the cycles in
$\widetilde{\mathcal{G}}$ are of constant sign, since there are no
flips in $\widetilde{\mathcal{G}}$. Thus, by our notation
$r_{i}=l_{i}$. Let us sum over the components:
\begin{equation}
    l(\mathrm{\bf{f}})=\sum_{i=1}^{\nu}l_{i}=\sum_{i=1}^{\nu}(B_{i}-V_{i}+1)=(B-
    \mathrm{F}(\textbf{f}))-V+\nu(\bf{f})
    \label{l(G)}
\end{equation}
Combining (\ref{quadratic form}) with (\ref{l(G)}), we get
(\ref{nodal domains via flips}).
\end{proof}

(\ref{nodal domains via flips}) is valid only for vectors $\bf f$
with no zero entries. In order to be able to handle a zero entry in
$\bf f$, we must perform a transformation on the graph. For a strong
nodal count, we simply delete all the zero vertices along with the
bonds connected to them from the graph, and then apply (\ref{nodal
domains via flips}) on the new graph (with the new Laplacian). For a
weak nodal count we replace each zero vertex by two vertices, one
positive and one negative (not connected to each other), and connect
them to all vertices which were connected to the original one. Now
we can apply (\ref{nodal domains via flips}), and get the desired
result.  Notice that this correction fails in the case of a zero
vertex whose neighbors are of the same sign (in this case an
artificial nodal domain is added). However, the situation above can
not occur for an eigenvector of a discrete Laplacian.  This way of
handling zero entries can be adapted for the following counting
methods as well and will not be repeated in the sequel.

Using (\ref{nodal domains via flips}), we can write some immediate
consequences:
\begin{eqnarray}
    \mathrm{F}(\textbf{f}_{n})+1-r\leq \nu_{n}\leq \mathrm{F}(\textbf{f}_{n})+1  \label{flips1}\\
     n-r-1\leq \mathrm{F}(\textbf{f}_{n})\leq n+r-1 \label{flips2}
\end{eqnarray}
(\ref{flips1}) results from the obvious fact that $0\leq l\leq r$,
while (\ref{flips2}) is a consequence of Courant's nodal domains
theorem and Berkolaiko's
theorem which states that $n-r\leq \nu_{n}$.\\
In order to make use of  (\ref{nodal domains via flips}), one must
compute $l(\bf{f})$ which is not given explicitly in terms of
$\textbf{f}$. Thus, it cannot be considered as an analytic
counting method, nor does it offer computational advantage (There
is no known efficient algorithm which counts all the cycles of
constant sign with respect to $\textbf{f}$). However, it offers a
useful analytical tool for deriving other results, and it makes a
useful connection between various quantities defined on the graph.

\subsection{Method \emph{II.} -- Partition function approach}

Foltin derived a partition function approach to counting nodal
domains of real functions in two dimensions \cite{Fo03}. It can be
adapted for graphs in the following way: Each vertex, $i$, is
assigned an auxiliary ``spin" variable $s_{i}$ where $s_{i}=\pm 1$
(a so called Ising-spin). Thus, given a certain function $\bf{f}$ on
the graph, each vertex is assigned with two ``spins"
 $s_i$ and $\widetilde{f}_i$. Let $\bf s$ denote  the
auxiliary spin vector: $\textbf{s}=(s_{1},s_{2},\ldots,s_{V})$.
Foltin introduced a \textit{weight} to each configuration of the
spins model. It assigns the value $1$ to configurations in which
all spins $s_{i}$ belonging to the same nodal domain (with respect
to $\bf{f}$) are parallel, while spins of different domains might
have different values. The weight reads:
\begin{equation}
w(\textbf{f},\textbf{s})=\prod_{i,j: \
C_{i,j}=1}\left[1-\frac{1+\widetilde{f}_{i}\widetilde{f}_{j}}{2}\cdot
\frac{1-s_{i}s_{j}}{2}\right] \ .
\end{equation}
It can be easily checked that this form satisfies the requirements
stated above: The weight can take the values one or zero. It is one
if and only if each factor in the product is equal to one. A certain
factor is one, in either one of the two cases: if
$\widetilde{f}_{i}\neq \widetilde{f}_{j}$ ($i,j$ belong to different
domains) - this allows the Ising-spins in different domains to be
independent of each other. The second case is if $i,j$ are in the
same domain, $\widetilde{f}_{i}=\widetilde{f}_{j}$ and the
corresponding Ising-spins are equal, $s_{i}=s_{j}$. Let us now sum
over all possible spin configurations $\{s_{i}\}$ to get the
partition function
\begin{equation}
Z(\textbf{f})\equiv\sum_{\{\textbf{s}\}}w(\textbf{f},\textbf{s})\ .
\label{eq:part}
\end{equation}
For the configurations whose weight has the value one, the spins
have equal signs over each nodal domain and different domains are
independent of each other. Hence, the total number of such
configurations is:
\begin{equation}
Z(\textbf{f})=2^{\nu(\textbf{f})}\ ,
\end{equation}
where $\nu(\textbf{f})$ is the number of nodal domains of the
vector \textbf{f}. The nodal domains count is:
\begin{equation}
\nu(\mathrm{\bf{f}})=\frac{1}{\ln2}\ln Z(\textbf{f})\approx1.44\ln
Z(\textbf{f}).
\end{equation}

The partition function approach provides an explicit formula for
the number of nodal domains, and therefore it belongs to the
analytic and not to the algorithmic counting methods. As a matter
of fact, it is highly inefficient for practical computations. It
involves running over all possible spin configurations
$\{s_{i}\}$, where $s_{i}=\pm1$. There are $2^{V}$ different
configurations, and as $V$ increases the efficiency deteriorates
rapidly.

The partition function approach can be used as a basis for the
derivation of some identities involving the graph and a vector $\bf
f$ defined on it. It is convenient to introduce the following
notations:
\[b\equiv\{i,j\}\]
\[\varphi_{b}\equiv\frac{1+\widetilde{f}_{i}\widetilde{f}_{j}}{2}\]
\[\sigma_{b}\equiv\frac{1-s_{i}s_{j}}{2}\]
Where $\widetilde{\bf f}$ and $\bf s$ are as before, and $b$ is an
undirected bond. We generalize the partition function by introducing
a new parameter $x$ into the definitions
\begin{eqnarray}
 w(\textbf{f},\textbf{s};x) &=& \prod_{b\in
\mathcal{B}}[1-\varphi_{b}\sigma_{b}x]  \label{partition function indicator form}\\
Z(\textbf{f};x) &\equiv&
\sum_{\{\textbf{s}\}}w(\textbf{f},\textbf{s};x)=\sum_{\{\textbf{s}\}}\prod_{b\in
\mathcal{B}}(1-\varphi_{b}\sigma_{b}x)   \label{partition function sum of indicator form}\\
Z(\textbf{f};1) &=& 2^{\nu(\textbf{f})}  \label{log of nodal
domains count}
\end{eqnarray}
where $x$ can assume any real or complex value. At $x=1$, the
generalized partition function is identical to (\ref{eq:part}).

Let us now perform the summation over all the vectors $\textbf{s}$,
and compute the coefficient of $x^{k}$. To get all the  contributing
terms we have to sum over all choices of $k$ brackets from
(\ref{partition function sum of indicator form}), in which $x$
appears. Non vanishing contributions occur whenever both
$\varphi_{b}$ and $\sigma_{b}$ are equal to one. Since we are only
summing over $\textbf{s}$, we only need to check when
$\sigma_{b}=1$. This happens if and only if the $\textbf{s}$ vector
has a flip on the bond $b$. Since we choose $k$ brackets (which is
equivalent to choosing $k$ bonds), we need to count how many
$\textbf{s}$ vectors have flips on all these $k$ bonds. The signs of
those $\textbf{s}$ vectors on bonds which are not contained in this
choice of $k$ bonds are irrelevant. If we observe the choice of $k$
bonds $(b_{1},\ldots,b_{k})$, we notice that each connected
component, within this choice, contributes a factor of 2, since the
symmetry of turning each plus to minus and vice versa, does not
change the flip properties. Using (\ref{eq:cycles}) we see that the
number of connected components with respect to the choice of $k$
bonds is: $Co(b_{1},\ldots,b_{k})=V-k+r(b_{1},\ldots,b_{k})$, where
$r(b_{1},\ldots,b_{k})$ is the number of independent cycles that are
contained in this choice. Finally we notice that a cycle of odd
length cannot have flips on all of its bonds, so we will not sum
over choices of $b_{1},\ldots,b_{k}$ which contain a cycle of odd
length. Thus, the sum over $\textbf{s}$ yields:
\begin{eqnarray}
Z(\textbf{f};x)&=&\sum_{k=0}^{B} \ \
\sideset{}{'}\sum_{b_{1},\ldots,b_{k}\in \mathcal{B}} \ \
\prod_{i=1}^{k}(\varphi_{b_{i}})2^{V-k+r(b_{1},\ldots,b_{k})}(-1)^{k}x^{k}
\nonumber\\
&= & {} 2^{V}\sum_{k=0}^{
B}\frac{(-1)^{k}}{2^{k}}\left[\sideset{}{'}\sum_{b_{1},\ldots,b_{k}\in
\mathcal{B}} \ \
\prod_{i=1}^{k}(\varphi_{b_{i}})2^{r(b_{1},\ldots,b_{k})}\right]x^{k}
\label{generalized partition function}
\end{eqnarray}
Where $\sideset{}{'}\sum_{b_{1},\ldots,b_{k}\in \mathcal{B}}$
stands for summation on all the possibilities to choose k bonds
$b_{1},\ldots,b_{k}\in \mathcal{B}$ such that the subgraph they
form do not contain an odd cycle. We can now derive some immediate
properties of the generalized partition function. To start, we
compute the leading four derivatives at $x=0$  to demonstrate the
counting techniques involved. Some more effort is required to
compute the higher derivatives.
\begin{eqnarray}
\hspace{1cm}  Z(\textbf{f};0) &=& \ 2^{V} \\
\hspace{1cm}
Z^{(1)}(\textbf{f};0)&=&-2^{V-1}\sideset{}{'}\sum_{b\in
\mathcal{B}}\varphi_{b}=-2^{V-1}(B-\mathrm{F}(\bf{f}))\\
  Z^{(2)}(\textbf{f};0)&=& \ 2! \
 2^{V-2}\sideset{}{'}\sum_{b_{1},b_{2}\in
\mathcal{B}}\ \prod_{i=1}^{2}(\varphi_{b_{i}})=2^{V-1}{B-\mathrm{F}(\bf{f}) \choose 2}\\
  Z^{(3)}(\textbf{f};0)&=&-3! \
 2^{V-3}\sideset{}{'}\sum_{b_{1},\ldots,b_{3}\in
\mathcal{B}}\ \prod_{i=1}^{3}(\varphi_{b_{i}})=-3! \
2^{V-3}\left[{B-\mathrm{F}(\bf{f}) \choose 3}-C_{3}\right]
\end{eqnarray}
\begin{eqnarray}
 \hspace{-1.8cm} Z^{(4)}(\textbf{f};0) &=& \ 4! \
2^{V-4}\sideset{}{'}\sum_{b_{1},\ldots,b_{4}\in \mathcal{B}}\
\prod_{i=1}^{4}(\varphi_{b_{i}}) \nonumber \\
\hspace{-1.8cm}&=&4! \
2^{V-4}\left\{2C_{4}+\left[{B-\mathrm{F}(\bf{f}) \choose
4}-C_{4}-\widetilde{C}_{3}\right]\right\}  \nonumber \\
 \hspace{-1.8cm} &=& 4! \ 2^{V-4}\left[{B-\mathrm{F}(\bf{f}) \choose
 4}+C_{4}-\widetilde{C}_{3}\right]
\end{eqnarray}
Where $C_{3}$ is the number of triangles of constant sign, $C_{4}$
is the number of cycles of length 4 of constant sign, and
$\widetilde{C}_{3}=C_{3}(B-\mathrm{F}(\textbf{f})-3)$ is the
number of choices of 4 non-flips bonds which contain a triangle.
In the evaluation of the function and its first three derivatives
we have used the identity $\forall n\leq3 \ \
r(b_{1},\ldots,b_{n})=0$ when $b_{1},\ldots,b_{n}$ contain no odd
cycles.

The partition function is a polynomial of degree less than or equal
to the number of bonds. For a graph $\mathcal{G}$ with $B$ bonds and
$Co$ connected components, the polynomial is of degree $B$ if and
only if $\mathcal{G}$ is bipartite and the function $\bf f$ is of
constant sign on each connected component. This follows from two
observations: first, a well known theorem in graph theory states
that a graph $\mathcal{G}$ is bipartite if and only if it contains
no cycles of odd length. Thus we can choose all the bonds in
$\mathcal{G}$ without having an odd cycle contained in this choice.
Second, unless $\bf f$ is of constant sign on each connected
component, we will encounter a flip and hence the multiplication
over all $\varphi_{b_{i}}$ will vanish. If $\mathcal{G}$ is
bipartite, the coefficient of $x^{B}$ is $2^{Co}$.

While the value of the polynomial at $x=1$ has an immediate
application through (\ref{log of nodal domains count}), we can
evaluate it for other values. Let us choose $\bf f$ to be a vector
of constant sign. This way $\varphi_{b_{i}}=1$ for all
$i=1,2,\ldots,B$. Hence (\ref{generalized partition function})
reduces to:
\begin{equation}
2^{V}\sum_{k=0}^{
B}\frac{(-1)^{k}}{2^{k}}\left(\sideset{}{'}\sum_{b_{1},\ldots,b_{k}\in
\mathcal{B}}2^{r(b_{1},\ldots,b_{k})}\right)x^{k}.
\label{partition function with constant f}
\end{equation}
On the other hand, (\ref{partition function sum of indicator form})
equals:
\begin{equation}
\sum_{\{\textbf{s}\}}\prod_{b\in \mathcal{B}}(1-\sigma_{b}x).
\label{partition function sum of indicator form with constant f}
\end{equation}
Choosing $x=-n+1$ where $n\in \mathbb{Z}$ and using the fact that
(\ref{partition function with constant f}) and (\ref{partition
function sum of indicator form with constant f}) are equal, we
get:
\begin{equation}
\frac{1}{2^{V}}\sum_{\{\textbf{s}\}}n^{\mathrm{F}(\bf
s)}=\sum_{k=0}^{
B}\frac{(-1)^{k}}{2^{k}}\left(\sideset{}{'}\sum_{b_{1},\ldots,b_{k}\in
\mathcal{B}}2^{r(b_{1},\ldots,b_{k})}\right)(-n+1)^{k}.
\end{equation}
For random vectors $\bf s$, uniformly distributed, the left hand
side can be interpreted as the average over this ensemble of the
quantity: $n^{\mathrm{F}(\bf s)}$. Let us choose two special
values for n. If we choose $n=-1$, the left hand side is just the
difference between the probability that a random vector $\bf s$ on
the graph has an even number of flips, and an odd number of flips.
The right hand side is:
\begin{equation}
\sum_{k=0}^{B}(-1)^{k}\sideset{}{'}\sum_{b_{1},\ldots,b_{k}\in
\mathcal{B}}2^{r(b_{1},\ldots,b_{k})}.
    \label{probability differences}
\end{equation}
For $n=2$, we get:
\begin{equation}
\frac{1}{2^{V}}\sum_{\{\textbf{s}\}}2^{\mathrm{F}(\textbf{s})}=\sum_{k=0}^{
B}\frac{1}{2^{k}}\sideset{}{'}\sum_{b_{1},\ldots,b_{k}\in
\mathcal{B}}2^{r(b_{1},\ldots,b_{k})}.
\end{equation}
An example of the use of the polynomial could be in proving that on
a tree, the probability of an even number of flips is equal to the
probability of an odd number of flips. We use Eq. (\ref{probability
differences}) and see that this difference of probabilities is equal
to:
$\sum_{k=0}^{B}(-1)^{k}{B \choose k}=0$.\\
Other possible identities we can derive are for the complete
graph, $K_{V}$. A function $\bf f$ of constant sign induces one
nodal domain on $K_{V}$, while any other function induces two
nodal domains. Therefore, for $x=1$:
\begin{eqnarray}
2^{V}\sum_{k=0}^{
B}\frac{(-1)^{k}}{2^{k}}\sideset{}{'}\sum_{b_{1},\ldots,b_{k}\in \mathcal{B}}2^{r(b_{1},\ldots,b_{k})}=2\\
\qquad \frac{2^{V}}{2^{V}-2} \sum_{\begin{subarray}
 \mathrm{\bf{f}}\in\{\pm1\}^V \ \\ \mathrm{\bf{f}}\neq\pm\overrightarrow{\bf{1}} \end{subarray}} \ \sum_{k=0}^{
B}\frac{(-1)^{k}}{2^{k}}\left[\sideset{}{'}\sum_{b_{1},\ldots,b_{k}\in
\mathcal{B}}\left(\prod_{i=1}^{k}\varphi_{b_{i}}(\mathrm{\bf{f})}\right)2^{r(b_{1},\ldots,b_{k})}\right]=4
\end{eqnarray}
Where $\varphi_{b_{i}}(\mathrm{\bf{f})}=\frac{1+f_{u}f_{v}}{2}$
for $b_{i} = (u,v)$.  Equivalently, running over all the functions
$\mathrm{\bf{f}}$ (including those of constant sign) we get:
\begin{eqnarray}
\quad 2^{V}\sum_{\mathrm{\bf{f}}\in\{\pm1\}^V}\sum_{k=0}^{
B}\frac{(-1)^{k}}{2^{k}}\left[\sideset{}{'}\sum_{b_{1},\ldots,b_{k}\in
\mathcal{B}}\left(\prod_{i=1}^{k}\varphi_{b_{i}}(\mathrm{\bf{f})}\right)2^{r(b_{1},\ldots,b_{k})}\right]=4(2^{V}-1)
\end{eqnarray}
\\


\subsection{Method \emph{III.} -- Breaking up the graph}

We begin again  by deleting all the flips from the graph
$\mathcal{G}$. This way we are left with a (possibly) disconnected
graph, $\widetilde{\mathcal{G}}$ in which each connected component
corresponds uniquely to a nodal domain in the original graph. The
connectivity matrix $\widetilde{C}$ and the discrete Laplacian
$\widetilde{L}$ of $\widetilde{\mathcal{G}}$ are given by
\begin{eqnarray}
\widetilde{C}_{ij} &=& C_{ij}\cdot
\frac{1+\widetilde{f}_{i}\widetilde{f}_{j}}{2}\\
\widetilde{L}_{ij} &=&
-\widetilde{C}_{ij}+\delta_{ij}\sum_{k=1}^{V}\widetilde{C}_{ik}
\end{eqnarray}
Where $\widetilde{\bf{f}}$ is the sign vector, and it is assumed
for the moment that none of the entries of ${\bf f}$ vanish.

We now make use of the theorem which states that the lowest
eigenvalue of the Laplacian is $0$ with multiplicity which equals
the number of connected components in the graph. Therefore,
finding the nodal domains count reduces to finding the
multiplicity of zero as an eigenvalue of $\widetilde{L}$. An
analytic counting formula can be derived by constructing the
characteristic polynomial of $\widetilde{L}$:
\begin{equation}
    \det(\lambda I_{V}-\widetilde{L})
    \label{det L tilde}
\end{equation}
The multiplicity of its $0$ eigenvalue provides the nodal domains
count:
\begin{equation}
\nu(\textbf{f})=\lim_{\lambda\rightarrow
0}\lambda\frac{d}{d\lambda}\ln\det(\lambda I_{V}-\widetilde{L}) \
.  \label{eq:logarithm derivative of det L tilde}
\end{equation}

\begin{figure}
{\includegraphics[0,0][292,193]{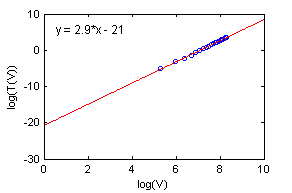}}
{\includegraphics[0,0][292,193]{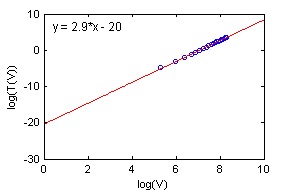}}
 \caption{The time it takes to compute the spectrum of $\widetilde{L}$ as a function of the number of vertices, for two different connectivity densities: $\frac{r}{V}=0.5$ in the upper figure and $\frac{r}{V}=5$ in the lower figure.  }  \label{fig:efficiency} \centering
\end{figure}

This method of counting, which provides the analytical expression
(\ref {eq:logarithm derivative of det L tilde}) for the nodal count,
is also the basis for a computational algorithm which turns out to
be very efficient. It relies on the efficiency of state of the art
algorithms to compute the spectrum (including multiplicity) of
sparse, real and symmetric matrices. To estimate the dependence of
the efficiency on the dimension $V$ of the graph, we have to
consider the costs of the various steps in the computation. The
construction of the matrix $\widetilde{L}$, takes $O(V^{2})$
operations, and storing the information requires $O(V^{2})$ memory
cells as well. It takes $O(V^\alpha)$ operations to find all its
eigenvalues where $\alpha\simeq2.3$ (and at worst case
$\alpha\simeq3$) \cite{DT99}. In figure \ref{fig:efficiency}, this
polynomial dependence is shown for graphs of two different
connectivity densities, with $\frac{r}{V}$ equals 0.5 and 5. In this
figure the logarithm of the time needed to find all eigenvalues of
$\widetilde{L}$ (defined for a random vector), is plotted against
the logarithm of the number of vertices. The slope which is the
exponent of the polynomial dependence is smaller than 3. The
eigenvalues in these two examples were attained using the Matlab
command \verb"eig". As will be shown below, there are more efficient
ways of finding the spectrum of sparse, real and symmetric matrices.
Thus, the efficiency stated above can be improved for graphs with
sparse Laplacians.
%
%

Finally, we would like to show that the present method can be
applied for counting nodal domains of functions defined on two
dimensional grids and that its efficiency is comparable to that of
the commonly used HK algorithm. Given a function $\textbf{f}$ on a
two dimensional domain, we have to compute its values on a
rectangular grid with $\sqrt{V}\times\sqrt{V}$ points. The HK
algorithm counts the nodal domains in $O(V)$ operations
\cite{HK76}. Using our method, we consider the rectangular grid as
a graph with $V$ vertices. Assuming for simplicity periodic
boundary conditions, the valency of all the vertices is 4. The
corresponding $L$ matrix is a $V\times V$ matrix which is sparse
(as long as $V\gg4$), and due to the periodic boundary conditions
it takes the explicit form:
\begin{equation}
L_{i,j}=4\delta_{i,j}-\delta_{i,j-1}-\delta_{i,j+1}-\delta_{i-V,j}-\delta_{i+V,j}
\ .
\end{equation}
Thus, storing $L$ takes only $O(V)$ memory cells, and constructing
$\widetilde{L}$ takes $O(V)$ operations. We mentioned above that
for a general real symmetric matrix, the number of operations
needed is $O(V^\alpha)$, where $\alpha\simeq2.3$ and at worst case
$\alpha\simeq3$. However, the sparse nature of $L$ significantly
simplifies the problem. The most well-known eigenvalue method for
sparse-real-symmetric matrices is the Lanczos method. In addition,
in recent years, new efficient algorithms were discovered for the
same problem. In \cite{SY}, it is proven that finding the
eigenvalues of a sparse symmetric matrix takes only $O(V)$
operations.
Combining the costs, we find that it takes our algorithm $O(V)$
operations in order to compute the nodal domains count, and
therefore it is comparable in efficiency to the HK
algorithm.\\
As mentioned earlier, the labeling algorithms also display linear
efficiency (\cite{DST92},\cite{FG96}). The labeling algorithms
have the advantage that they are simpler in a sense, and that they
are implemented quite easily as computer programs. In addition,
the labeling algorithms maintain their linear efficiency even for
graphs with dense Laplacians.  It is worth mentioning, however,
that our algorithm has the advantage that it provides an analytic
expression of the nodal domains count (\ref{eq:logarithm
derivative of det L tilde}).

\subsection{Method \emph{IV.} -- A geometric point of view}

The counting method proposed here uses a geometric point of view
which starts by considering the $V$ dimensional Euclidean space, and
dividing it into  $2^V$ \emph{sectors} using the following
construction. Consider the $2^V$ vectors
$\textbf{e}^{(\alpha)}=(e^{(\alpha)}_{1},e^{(\alpha)}_{2},\ldots,e^{(\alpha)}_{V})$
where $e^{(\alpha)}_{i}=\{1,-1\}$, $\alpha=1,2,\ldots,2^{V}$. A
vector $\textbf{x}\in \mathbb{R}^V$ is in the sector $\alpha$ if
$\textbf{x}\cdot\textbf{e}^{(\alpha)} = \sum_{i=1}^{V}{\mid
x_{i}\mid}$. In two dimensions, the sectors are the standard
quadrants. We shall refer to the vectors $\textbf{e}^{(\alpha)}$ as
the \textit{indicators}.

Given a graph $\mathcal{G}$ with $V$ vertices, we partition the
$2^{V}$ indicators into disjoint sets:
$\gamma_{n}=\{\textbf{e}^{(\alpha)}:\nu(\textbf{e}^{(\alpha)})=n\}$
where $\nu(\textbf{e}^{(\alpha)})$ denotes the nodal domains count
of the indicator $\textbf{e}^{(\alpha)}$ with respect to
$\mathcal{G}$. As shown before $
\rm{max}\{n\,|\,\gamma_{n}\neq\emptyset\}\leq V-\chi+2$, where
$\chi$ is the chromatic number of $\mathcal{G}$, and also some of
the $\gamma_{i}^{'}s$ might be empty.

Let $\textbf{f}$ be a vector with non-zero entries defined on the
vertex set of $\mathcal{G}$. Then, the main observation is that
$\nu(\textbf{f})=n$ if and only if:
\begin{equation}
\sum_{\textbf{e}^{\alpha}\in
\gamma_{n}}\hat{\delta}\left(\left(\textbf{e}^{\alpha},
\textbf{f}\right)-\sum_{i=1}^{V}\mid \textbf{f}_{i}\mid \right)\ =\
1\ ,
\end{equation}
where,
\begin{equation}
\hat{\delta}(x)=\lim_{\epsilon\rightarrow0}\frac{\epsilon}{x}\sin{\frac{x}{\epsilon}}=\left\{\begin{array}{ll}
1, & \textrm{if $x=0$}\\ 0, & \textrm{if $x\neq0$}
\end{array} \right. \ .
\end{equation}
and $\left(\textbf{e}^{\alpha}, \textbf{f}\right)$ is the dot
product of $\textbf{e}^{\alpha}$ and $\textbf{f}$. In other words,
by finding the sector to which $\textbf{f}$ belongs and knowing from
a preliminary computation the number of nodal domains in each
sector, one obtains the desired nodal count. Thus, the present
method requires a preliminary computation in which the sectors are
partitioned into equi-nodal sets $\gamma_{n}$. This should be
carried out once for any graph. Therefore the method is useful when
the nodal counts of many vectors is required. In several
applications,  one is given a vector field (of unit norm for
simplicity) $\textbf{f}\in \mathbb{S}^{V-1}$ which is distributed on
the $(V-1)$-sphere with a given probability distribution
$p(\textbf{f})$, and one is asked to compute the distribution of
nodal counts,
\begin{equation}
P(n)=\int_{S^{V-1}}p(\textbf{f})\,\hat{\delta}\left(\nu(\textbf{f})-n\right)\,d^{V-1}\textbf{f}
\ .
\end{equation}
In such cases, the preliminary task of computing the equi-nodal pays
off, and one obtains the following analytic expression for the
distribution of the nodal counts.
\begin{eqnarray}
\lefteqn{P(n)=\int_{S^{V-1}}p(\textbf{f})\,d^{V-1}\textbf{f}\sum_{\textbf{e}^{\alpha}\in
\gamma_{n}}\hat{\delta}\left(\left(\textbf{e}^{\alpha},
\textbf{f}\right)-\sum_{i=1}^{V}| \textbf{f}_{i}|\right)={}} \nonumber\\
& & {} \int_{\mathbb{R}^{V}}\tilde{p}(\textbf{f})\,\delta\left(1-|
\textbf{f}|^{2}\right)\,d^{V}\textbf{f}\sum_{\textbf{e}^{\alpha}\in
\gamma_{n}}\hat{\delta}\left(\left(\textbf{e}^{\alpha},
\textbf{f}\right)-\sum_{i=1}^{V}| \textbf{f}_{i}|\right) \label{P(n)
1}
\end{eqnarray}
where:
$\tilde{p}(\textbf{f})=p(\frac{\textbf{f}}{|\textbf{f}|}) \cdot 2|\textbf{f}|$.\\
(\ref{P(n) 1}) can also be formulated as:
\begin{eqnarray}
\lefteqn{P(n)=\sum_{\beta=1}^{2^{V}}\int_{f_{i}\geq0}\tilde{p}(\textbf{f}^{\beta})\,\delta(1-|
\textbf{f}|^{2})\,d^{V}\textbf{f}\sum_{\textbf{e}^{\alpha}\in
\gamma_{n}}\hat{\delta}\left(\left(\textbf{e}^{\alpha},
\textbf{f}^{\beta}\right)-\sum_{i=1}^{V}f_{i}\right)={}}
\nonumber\\
& & {} \sum_{\textbf{e}^{\beta}\in
\gamma_{n}}\int^{+}\tilde{p}(\textbf{f}^{\beta})\,\delta(1-|
\textbf{f}|^{2})\,d^{V}\textbf{f}\sum_{\textbf{e}^{\alpha}\in
\gamma_{n}}\hat{\delta}\left(\left(\textbf{e}^{\alpha},
\textbf{f}^{\beta}\right)-\sum_{i=1}^{V}f_{i}\right) \label{P(n) 2}
\end{eqnarray}
Where $\int^{+}=\int_{f_{i}\geq0}$ means integration on the first
sector (the vectors with all entries positive) and
$\textbf{f}^{\mu}=(f_{1}e_{1}^{\mu},f_{2}e_{2}^{\mu},\ldots,f_{V}e_{V}^{\mu})$.

Equations (\ref{P(n) 1}) and (\ref{P(n) 2}) are the general
equations governing the nodal domains count distribution. In order
to make further progress, we need to specify the distribution from
which $\bf f$ is taken. This means that we need to specify
$\tilde{p}(\textbf{f})$ in (\ref{P(n) 1}) for example. Let us
discuss two examples: \\
\noindent  \emph{A uniform distribution over the $V-1$ dimensional
sphere:} In this case, we can solve Equation  (\ref{P(n) 1}) and get
that $P(n)=\frac{|\gamma_{n}|}{2^{V}}$. Note that for a tree, we can
solve this problem by other means. Using (\ref{nodal domains via
flips}), we see that for a tree, the number of nodal domains is
equal to the number of flips plus one. Since $\bf f$ is taken from
the uniform distribution, then the probability of a flip is half.
The number of flips in a vector $\bf f$ is thus a binomial variable:
$\mathrm{F}(\textbf{f})\sim Binomial(N,p)$ with $N=V-1$ is the
number of bonds, and $p=\frac{1}{2}$. For large enough $V$ this
approaches the Gaussian distribution: $\mathrm{F}(\textbf{f})\sim
Gaussian(\mu,\sigma^{2})$ with $\mu=\frac{V-1}{2}$ and
$\sigma^{2}=\frac{V-1}{4}$. From this result we can infer that:
\begin{eqnarray}
P(n)\approx\frac{2}{\sqrt{2\pi
(V-1)}}\exp{(\frac{-2(n-\frac{V+1}{2})^{2}}{V-1})} \\
|\gamma_{n}|\approx\frac{2^{V+1}}{\sqrt{2\pi
(V-1)}}\exp{(\frac{-2(n-\frac{V+1}{2})^{2}}{V-1})} \label{Gaussian
approx. to binomial tree}
\end{eqnarray}
For the other extreme, the complete graph, $K_{V}$, the only
possible nodal domains counts are one and two \cite{Or07}. The
vectors which yield a nodal domains count of one are vectors of
constant sign. All other vectors yield a nodal domains count of
two. Indeed, using (\ref{P(n) 1}) or (\ref{P(n) 2}) it is easy to
be convinced that for the complete graph, $\gamma_{1}=2$ while
$\gamma_{2}=2^{V}-2$. All other $\gamma_{n}$'s are empty. \\
\noindent \emph{Micro-canonical ensemble:} In this case the vectors
$\bf f$ are uniformly distributed on the energy shell, where we can
also define a measurement tolerance factor, $\Delta$:
\begin{equation}
p_{E}(\textbf{f})=\frac{\delta\left(E-| \left(\textbf{f}, L
\textbf{f}\right)-\Delta|\right)\,\delta\left(1-| \textbf{f}
|^{2}\right)}{\int_{S^{V-1}}d^{V-1}\textbf{f}\,\,\delta\left(E-|
\left(\textbf{f}, L\textbf{f}\right)-\Delta|\right)}
\end{equation}
In order to make use of this ensemble, further work must be done,
for example, a natural way to order the functions of the ensemble.

\section{The resolution of isospectrality}
There are several known methods to construct isospectral yet
different graphs.  A review of this problem for discrete graphs can
be found in \cite{Brooks}. The conditions under which the spectral
inversion of quantum graphs is unique were studied previously. In
\cite {vonBelow,Carlson99} it was shown that in general, the
spectrum
 does not determine uniquely the length of the bonds and their
connectivity. However, it was shown in \cite{gutkinus} that
quantum graphs whose bond lengths are rationally independent ``can
be heard" - that is - their spectra determine uniquely their
connectivity matrices and their bond lengths. This fact follows
from the existence of an exact trace formula for quantum graphs
\cite {Roth,KS}.  Thus, isospectral pairs of non congruent graphs,
must have rationally dependent bond lengths.
An example of a pair of metrically distinct graphs which share the
same spectrum was already discussed in \cite{gutkinus}.

\begin{figure}[h] \centering
\scalebox{.1}{\includegraphics[0,0][3724, 2232]{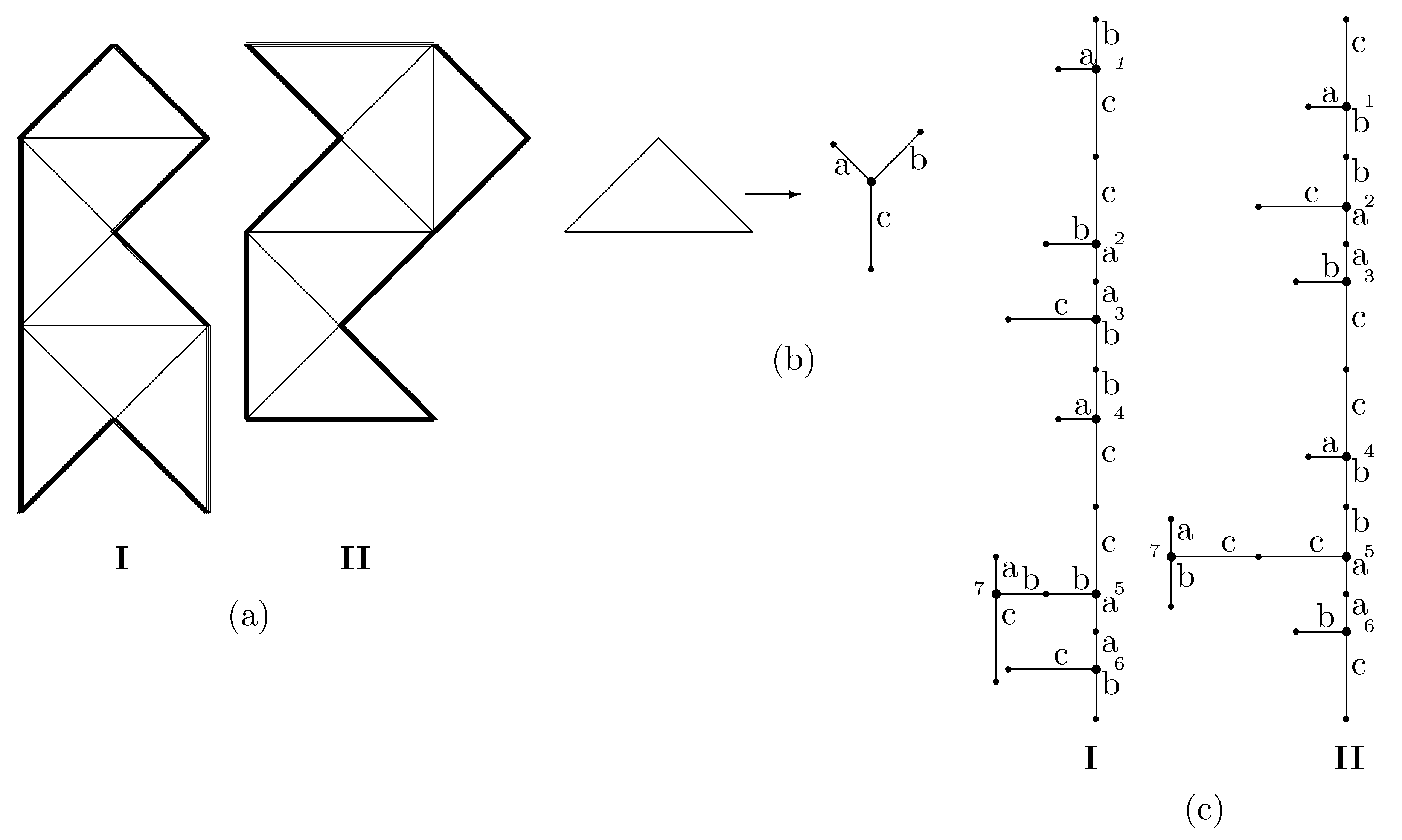}}
 \caption{(a) Planar isospectral domains of the $7_3$ type.
(b) Reducing the building block to a 3-star. (c) The resulting
isospectral quantum graphs.}  \label{fig:7-3}
\end{figure}

The main method of construction of isospectral pairs is due to
Sunada \cite{Sunada}.  This method enabled the construction of the
first pair of planar isospectral domains in $\mathbb{R}^{2}$
\cite{Gordon} which gave a negative answer to Kac's original
question: `Can one hear the shape of a drum?' \cite{kac}. Later,
it was shown that all the known isospectral domains in
$\mathbb{R}^{2}$ \cite {Buser,Okada} which were also constructed
using the Sunada method have corresponding isospectral pairs of
quantum graphs \cite{Tashkent}. An example of this correspondence
is shown in figure \ref{fig:7-3}.
As mentioned in the introduction, it is conjectured
\cite{GSS05,BSS06} that nodal domains sequences resolve between
isospectral domains. For flat tori in 4-d, this was proven
\cite{BKP07}.  We present here three additional known results for
the validity of the conjecture for graphs.

 The first result is for the quantum graphs shown in figure
 \ref{fig:7-3}(c).  Both graphs of this isospectral pair are
 tree graphs and therefore have the same metric nodal count
 $\mu_n=n$ \cite{Sch06}.  This demonstrates the
 need to use the discrete nodal count in order to resolve
isospectrality in this case.  Indeed numerical examination of this
case shows that for the first 6600 eigenfunctions there is a
different discrete nodal count for $\simeq$ 19 $\%$ of the
eigenfunctions.  Similar numerical results exist for two other pairs
of isospectral graphs that are constructed from the isospectral
domains in \cite {Buser,Okada}.  The exact results are described in
\cite{BSS06}.

\begin{figure}[h]
  \centering
 \scalebox{0.62}{\includegraphics[0,0][234,124]{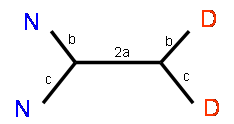}}
  \scalebox{0.55}{\includegraphics[0,0][332,129]{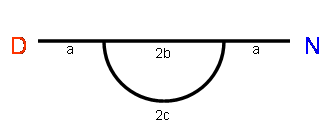}}
 \\   \hspace{-0.7cm} I \hspace{5.3cm}  II
  \caption{The isospectral pair with boundary conditions. D stands
for Dirichlet and N for Neumann.  The bonds' lengths are determined
by the parameters a,b,c}
  \label{fig:isospectral pair+notations}
\end{figure}

  Another result is also in the field of
quantum graphs \cite{BSS06}. The graphs in figure
\ref{fig:isospectral pair+notations} are the simplest isospectral
pair of quantum graphs known so far.
The simplicity of these graphs enables the comparison between the
nodal counts of both graphs.  It was proved that the nodal count
is different between these graphs for half of the spectrum.  This
result was proved separately for the discrete count and for the
metric count.  The proof does not contain an explicit formula for
the nodal count but rather deals with the difference of the nodal
count between the graphs averaged over the whole spectrum.

Examining the nodal sequences for the graph $II$ for various values
of the length parameters $a,b,c$, we observed that the formula
\begin{equation}
\mu^{II}_n = n - \frac{1}{2} -
\frac{1}{2}(-1)^{\lfloor\frac{b+c}{a+b+c}n\rfloor}
\label{eq:nodal_count_formula}
\end{equation}
reproduces the entire data set without any flaw \footnote{This
result was obtained with A. Aronovitch.} . Assuming it is correct
(which is not yet proved rigorously) we first see that it provides
an easy explanation for the previously discussed result regarding
the resolution of isospectrality for this pair. For rationally
independent values for the parameters a,b,c one gets that
$\mu^{II}_n \neq n$ for half of the spectrum. Combining this with
$\mu^{I}_n = n$ (since graph $I$ is a tree) we see again that for
half of the spectrum the nodal domain sequences are different.
Expression (\ref {eq:nodal_count_formula}) is a periodic function of
$n$ with period proportional to the length of the only loop orbit on
the graph (the length is measured in units of the graph's total
length). It can be expanded and brought to a form which is similar
in structure to a trace formula where the length of this orbit and
its repetitions are the oscillation frequencies. A similar trace
formula for the nodal counts of the Laplacian eigenfunctions on
surfaces of revolution was recently derived \cite{GPS06}.

Finally, we direct our attention to discrete Laplacians.  It was
recently shown \cite{Or07} that if $\mathcal{G}$ and $\mathcal{H}$
are two isospectral graphs where one of them is bipartite and the
other one is not, then their nodal domains count will differ.
Without loss of generality, let $\mathcal{G}$ be a bipartite graph
and $\mathcal{H}$ a non-bipartite one, then for the eigenvector of
the largest eigenvalue, the nodal domains count are different: for
$\mathcal{G}$, $\nu_{_V}=V$, while for $\mathcal{H}$, $\nu_{_V}<V$.
The proof of this theorem is based on another interesting result
derived in \cite{Or07}: Denote by $\mathrm{\bf{f}}_{V}$ the
eigenvector corresponding to the largest eigenvalue of the Laplacian
of a connected graph $\mathcal{G}$. Then
$\nu(\mathrm{\bf{f}}_{V})=\nu_{\mathcal{G}}=V$, if and only if
$\mathcal{G}$ is bipartite. Figure 4. illustrates this result.

\begin{figure}[!h]
  \centering
{\includegraphics[0 ,0][164,94]{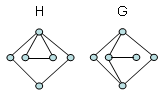}}
{\includegraphics[0,0][292,193]{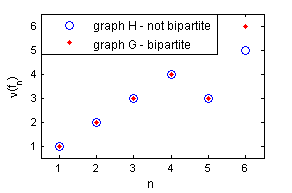}}
 \caption{The upper figure presents a pair of isospectral graphs taken from
\cite{HaSp04}. Graph $G$, on the right is bipartite, whereas graph
$H$, on the left, is not. The lower figure presents the nodal
domains count, $\nu(\textbf{f}_{n})$ \textit{vs.} the index $n$.
}
 \end{figure}

\section{Summary and open questions}
In spite of the  progress achieved recently in the study of nodal
domains on graphs, there are several outstanding open problems which
call for further study.  We list here a few examples.

Of fundamental importance is to find out whether there  exists a
``trace formula" for the nodal count sequence of graphs, similar to
the one derived in \cite {GPS06} for surfaces of revolution. The
closest we reached this goal is for the graph $II$ in the previous
section,  where (\ref {eq:nodal_count_formula}) could be expanded in
a Fourier series. However (\ref{eq:nodal_count_formula}) was deduced
numerically but not proved. Once a \emph{nodal} trace formula is
available, it could be compared to the \emph{spectral} trace formula
\cite{KS} and might show the way to prove or negate the conjecture
that counting nodal domains resolves isospectrality
\cite{GSS05,BSS06}.

The conjecture mentioned above can be addressed from a different
angle.  One may study the various systematic ways to construct
isospectral pairs and investigate the relations between the
construction method and the nodal count sequence of the resulting
graphs.  Such an approach worked successfully for the isospectral
graphs presented in figure \ref{fig:isospectral pair+notations}
\cite{BSS06}.

Another open question which naturally arises in the present context:
Can one find graphs whose Laplacians have different spectra but the
nodal count sequences are the same? A positive answer is provided
for tree graphs \cite{Sch06}. Are there other less trivial examples
of ``isonodal"  yet not isospectral domains?

It follows from Berkolaiko's theorem \cite{Be06} that the number
of nodal domains (both metric and discrete) of the $n^{th}$
eigenfunction is bounded in the interval $[n-r,n]$. We can thus
investigate the probability to have a nodal count  $\nu_n =
n-\tilde{r}$ (for $0\leq\tilde{r}\leq r$). This probability, which
is defined with respect to a given ensemble of graphs, is denoted
by $P(\tilde{r})$. It is defined for discrete graph Laplacians as:
\begin{eqnarray}
P(\tilde{r}) \equiv\frac{1}{V}\langle \# \left\{ \ 1\leq n\leq V:\
\nu_n=n-\tilde{r} \right\}\rangle \ .
\end{eqnarray}
The corresponding quantity for  metric Laplacians is:
\begin{eqnarray}
 \nonumber N(K) & \equiv & \langle \# \left\{n : k_n\leq K \right\}\rangle \\
  \nonumber P(\tilde{r} ;K) & \equiv & \frac{1}{N(K)}\ \langle \# \left\{ \
n\le N(K)\ :\ \mu_n=n-\tilde{r} \right\}\rangle \\
 P(\tilde{r})& \equiv &\lim_{K\rightarrow\infty} P(\tilde{r}; K)
\end{eqnarray}
Here, $\langle\ \ \ \rangle$ stands for the expectation with
respect to the ensemble.  New questions arise from the
investigation of the relation between the connectivity of the
graph and the nodal distribution $P(\tilde{r})$. Can one use the
information stored in $P(\tilde{r})$ to gain information on the
graphs e.g., the mean and the variance of the valency (degree)
distribution of the vertices in the graphs?

 Many of the results we have presented, have analogues in
Riemannian manifolds (which in most cases, were discovered earlier)
- for example, Courant's theorem was originally formulated for
manifolds. One can search for other analogues, and a good example is
the Courant-Herrmann Conjecture (CHC).  For manifolds the CHC states
that any linear combination of the first $n$ eigenfunctions divide
the domain, by means of its nodes, into no more than $n$ nodal
domains. Gladwell and Zhu \cite{GZ03} have shown that in general
there is no discrete counterpart to the CHC. However, we can still
ask for which classes of graphs does the CHC hold?

\section {Acknowledgments}
The authors would like to thank L. Friedlander for stimulating
discussions and suggestions of interesting open problems. It is a
pleasure to acknowledge A. Aronovitch and Y. Elon for their helpful
comments and well thought of critical remarks. We would like to
thank the organizers of the AGA program of the I. Newton Institute,
and in particular  P. Kuchment for his support and encouragement.
The work was supported by the Minerva Center for non-linear Physics
and the Einstein (Minerva) Center at the Weizmann Institute, and by
grants from the EPSRC (grant 531174), GIF (grant I-808-228.14/2003),
and BSF (grant 2006065).


\end{document}